\documentstyle[psfig]{mn}
\def\H0{{\it H}$_0$~}
\def\Ms{{\it M}$_\odot$}

\def\q0{{\it q}$_0$~}
\def\kmps{km~s$^{-1}$}
\def\ergps{erg~s$^{-1}$}

\def\min{$^{\prime}$}

\def\nH{$N_{\rm H}$~} 
\def\psqcm{cm$^{-2}$}
\def\ergpspsqcm{erg~cm$^{-2}$~s$^{-1}$}

\def\ll{\leftline}
\def\Zs{$Z_{\odot}$}
\def\Lx{$L_{\rm X}$}

\def\cps{ct s$^{-1}$}
\def\pcmcu{cm$^{-3}$}

\title[X-ray spectrum of NGC4388]
{ASCA PV observations of the Seyfert 2 galaxy NGC 4388: the obscured nucleus and its X-ray emission}
\author[K. Iwasawa {\it et al.}]
{\parbox[]{6.5in}{K. Iwasawa$^1$, A.C. Fabian$^1$, S. Ueno$^2$, 
H. Awaki$^2$, Y. Fukazawa$^3$, K.~Matsushita$^3$ and K. Makishima$^3$}\\
\\
1: Institute of Astronomy, Madingley Road, Cambridge CB3 0HA \\
2: Department of Physics, Kyoto University, Sakyo-ku, Kyoto 606-01, Japan\\
3: Department of Physics, the University of Tokyo, Hongo, Bunkyo-ku, 
Tokyo 113, Japan}

\date{}

\begin{document}

\maketitle

\begin{abstract}
We present results on the Seyfert 2 galaxy NGC4388 in the Virgo cluster
observed with {\sl ASCA} during its performance verification (PV) phase. The
0.5--10 keV X-ray spectrum consists of multiple components; (1) a
continuum component heavily absorbed by a column density \nH$\approx 4 \times
10^{23}$\psqcm ~above 3 keV; (2) a strong 6.4 keV line (equivalent width
$EW\sim 500$ eV); (3) a weak flat continuum between 1 and 3 keV; and (4)
excess soft X-ray emission below 1 keV. The detection of strong
absorption for the hard X-ray component is firm evidence for an obscured
active nucleus in this Seyfert 2 galaxy. The absorption corrected X-ray
luminosity is about $2\times 10^{42}$\ergps. This is the first time that
the fluorescent iron-K line has been detected in this object; the large
EW is a common property of classical Seyfert 2 nuclei. The flat spectrum
in the intermediate energy range may be a scattered continuum from
the central source. The soft X-ray emission below 1 keV can
be thermal emission from a temperature $kT\simeq 0.5$ keV, consistent
with the spatially extended emission observed by {\sl ROSAT} HRI.
However, the low abundance ($Z\sim 0.05$\Zs) and high mass flow rate
required for the thermal model and an iron-K line stronger than expected from 
the obscuring torus model are puzzling. An alternative consistent solution 
can be obtained if the central source was a hundred times more luminous
over than a thousand years ago. All the X-ray emission below 3 keV is then 
scattered radiation. 

\end{abstract}


\section{Introduction}

\begin{figure*}
\centerline{\psfig{figure=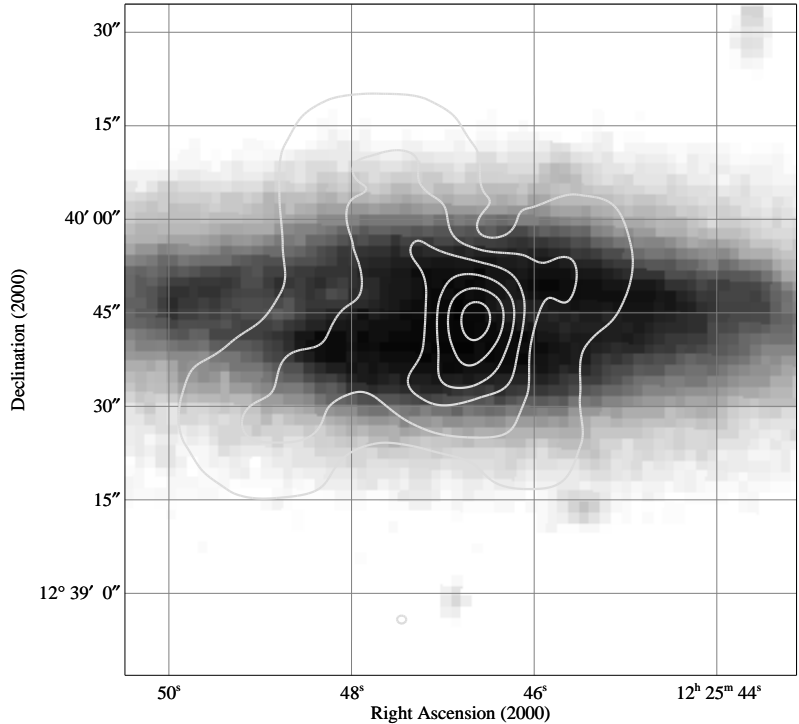,width=0.6\textwidth,angle=0}}
\caption{The {\sl ROSAT} HRI image (Matt et al 1994) overlaid 
on the blue optical band image taken from the Palomer 48-inch 
Schmidt Digitized Sky Survey.}
\end{figure*}

NGC4388 is an active galaxy situated near the core of the Virgo cluster.
The spiral stellar disk (SB(s)b pec; Phillips \& Malin 1982) is viewed
nearly edge-on (inclination $\sim 72^{\circ}$) with a dust lane across
near the nucleus. The distance of the galaxy is assumed to be 19.7 Mpc
(Sandage \& Tammann 1984) throughout this paper, giving an angular scale
of 95.5 pc arcsec$^{-1}$. The blue and far infrared luminosities are
$L_{\rm B}\simeq 6.4\times 10^{43}$\ergps ~and $L_{\rm FIR}\simeq 2.7\times 
10^{43}$\ergps, assuming this distance (David, Jones \& Forman 1992).

Optical spectra of the nucleus show emission lines characteristic of a
Seyfert type 2 galaxy (Phillips \& Malin 1982; Phillips, Charles \&
Baldwin 1983). An anisotropic ionization field is indicated by the cones
of highly-ionized line emission extending above and below the stellar
disk (Colina et al 1987; Pogge 1988). Gas emitting [OIII]$\lambda 5007$
extends as far as 50 arcsec ($\sim$5kpc) from the nucleus (Pogge 1988).
The ionization state of the gas is best described as due to
photoionization by a powerful nonthermal source (Pogge 1988; Colina
1992). Collimation of the nuclear radiation is strongly supported by the
VLA images (Stone, Wilson \& Ward 1988; Hummel \& Saikia 1991).

The detection of a weak broad wing to H$\alpha$ (FWZI $\simeq 6000$
\kmps) reported by Filippenko \& Sargent (1985) suggests the presence of
a Seyfert 1 nucleus obscured from our direct view. Shields \& Filippenko
(1988, 1996) also find off-nuclear broad H$\alpha$ emission (FWZI $\simeq$4000
\kmps ) arising within the biconical high-ionization regions observed by
Pogge (1988). A plausible origin for these properties is the scattering
of line emission from a hidden Broad Line Region.

A strong hard X-ray source in NGC4388 has been detected by the Birmingham
X-ray coded mask telescope (SL2~XRT: Hanson et al 1990), the {\sl Ginga} LAC
(Takano \& Koyama 1991) and SIGMA (Lebrun et al 1992), supporting the
hypothesis of a powerful hidden nucleus. The results suggest the presence
of a power-law X-ray source of photon index, $\Gamma\approx 1.5$,
absorbed by a column density \nH $\sim$ a few times $10^{23}$\psqcm. The 
absorption-corrected, 2--10 keV luminosity \Lx =$2\times 10^{42}$
\ergps, is typical for a low luminosity Seyfert 1 galaxy.

The origin of weak soft X-ray emission seen from the source is, on the
other hand, controversial The soft X-ray luminosity, $L_{\rm 0.5-3 keV}
= 1.5\times 10^{40}$\ergps, first detected by the {\sl Einstein
Observatory} (Forman et al 1979), is $\sim$1 per cent of the
absorption-corrected luminosity of the hard X-ray source and consistent
with the possibility that the soft X-rays are scattered nuclear
radiation. However, the {\sl ROSAT} HRI image of NGC4388 shows the 0.1--2.4
keV emission to be clearly extended with a radius of 45 arcsec and
luminosity \Lx $\simeq 3\times 10^{40}$\ergps (Matt et al 1994; see also 
Fig. 1). The
radial profile shows that half of the soft X-ray emission originate
beyond the central 15 arcsec. Matt et al (1994) claim that the large extent is
evidence against a scattering origin for the soft X-ray emission, since
the X-ray luminosity of the central source implied by the hard X-ray
observations is insufficient to photoionize the necessary column density.
They propose as alternative explanations either thermal emission from a
starburst or a collection of discrete sources. The {\sl ROSAT} PSPC spectrum
shows a steep spectral slope, $\Gamma =$2--3 (Rush \& Malkan 1996) which
may be consistent with this interpretation. However, the spectral
resolution of these instruments does not provide any conclusive evidence.

We report here the first {\sl ASCA} results on NGC4388 (see Tanaka, Inoue \&
Holt 1994 for a brief discussion of the capabilities of {\sl ASCA}). The
superior spectral resolution of {\sl ASCA} over the 0.5--10 keV band enables us
to identify several different components in the spectrum.

\section{The ASCA observations and data reduction}
\subsection{Observations}

NGC4388 was observed in two different pointings at adjacent elliptical
galaxies in the Virgo cluster, M86 (NGC4406; {\sl ASCA} results of this galaxy
appear in Awaki et al 1994 and Matsushita et al 1994) and M84 (NGC4374)
during the {\sl ASCA} performance verification (PV) phase. These observations
were carried out in series between 1993 July 3 and 4, each with an
effective exposure time of about 20 ks. NGC4388 was off axis by 18 and 10
arcmin, respectively. In the first observation, NGC4388 was outside the
field of view of the Solid state Imaging spectrometer (SIS) detector so
no SIS data was obtained. Data reduction was made using the {\sl ASCA} standard
softwares, XSELECT and FTOOLS. One dataset of effective exposure time 17
ks for both SIS (S0 and S1), and two datasets of each 20 ks were obtained
for both Gas Imaging Spectrometers (GIS: G2 and G3). 

\subsection{The image}

The full band (0.7--10 keV) GIS image from the first observation is shown
in Fig. 2. The two elliptical galaxies, M86 and M84 are present in the
field of view as well as NGC4388. Distortion of the X-ray image of
NGC4388 is due to the off-axis response of the X-ray Telescope (XRT) and
is consistent with the point spread function at that position
(the elongation of the image aligning with the galaxy major axis is
a coincidence). The SIS image obtained at a small off-axis position
does not show evidence for significant extended emission within the
spatial resolution of ASCA.
Fig. 3 indicates that NGC4388 is the only strong source in the energy
band above 3 keV but is faint in the softer band. Significant X-ray
emission at the position of NGC4388 is, however, detected both in the GIS
and SIS in the energy band below 3 keV as an excess above the diffuse
emission spreading out from the M86 direction.

\begin{figure*}
\centerline{\psfig{figure=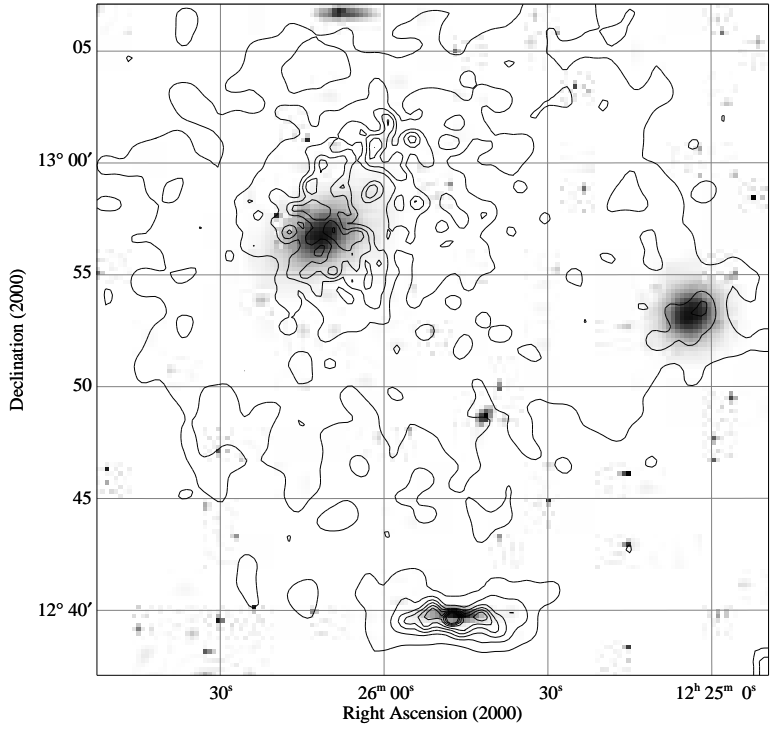,width=0.6\textwidth,angle=0}}
\caption{Full-band (0.7--10 keV) GIS image from the first observation, 
overlaid on the digital sky survey image in the optical band. 
X-ray emission was detected from three galaxies in the field of view; 
M86 (middle), M84 (right),
and NGC4388 (below). 
}
\end{figure*}

\begin{figure*}
\centerline{\psfig{figure=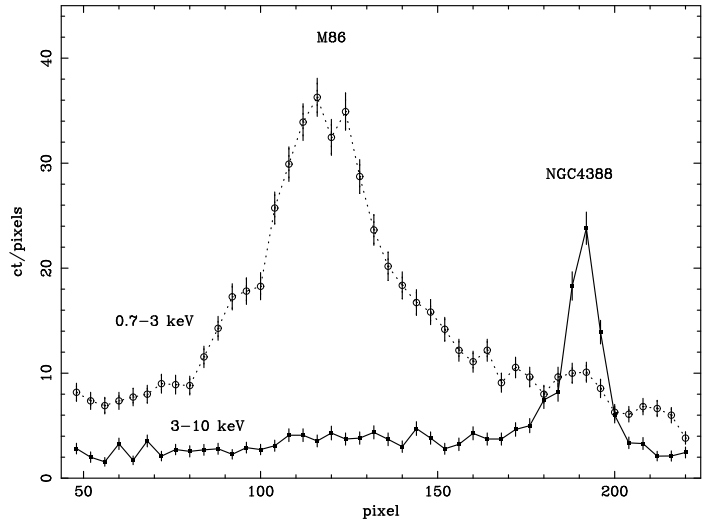,width=0.6\textwidth,angle=270}}
\caption{Projected images across M86 and NGC4388 in two energy bands; 
open circles with a dotted line: 0.7--3 keV; and filled squares with
a solid line: 3--10 keV. 1 pixel corresponds to 0.25 arcmin.}
\end{figure*}

\subsection{Spectral data and diffuse background}

Energy spectra were extracted from an elliptical-shaped region with the
major axis aligned with the elongation of the image (3 $\times$ 1.6
arcmin for the SIS; 5 $\times $3 arcmin for the GIS).

Background subtraction must be done carefully, both because of the
weakness of NGC4388 itself and the complex underlying diffuse emission in
this busy field. Since NGC4388 lies close to the core of the Virgo
cluster there is significant emission from the intra-cluster medium
(ICM). The temperature of the ICM, $kT\approx 2.5$ keV, has been measured
with {\sl Ginga} (Koyama, Takano \& Tawara 1991) and {\sl ROSAT} (B\"ohringer et al
1994). In addition, there may be some extended emission from
M86\footnote{Awaki et al (1994) give a temperature of $kT = 0.79\pm
0.01$ keV and abundance $Z = 0.45^{+0.13}_{-0.08}$\Zs for the total X-ray
emission of M86 after fitting with a Raymond-Smith thermal spectrum model.}
which makes the background around NGC4388 complicated.

We have checked the background data by dividing it into four distinct
regions. The data taken from the far side (SW of NGC4388) of M86 and M84
are weaker than in the other directions and show a different spectral
shape. This implies that the diffuse emission has a gradient in intensity
and spectrum across the source-extracted region of NGC4388. We thus took
background data from a circular, annulus of width 5 arcmin for the SIS
and a circular or an elliptical annulus (depending on off-axis angle)
with a radius of 6--8 arcmin for the GIS. The background used here should
be a good approximation of the true background, since taking data from
the region surrounding the source should smear the gradient out.

The background-subtracted count rate in each detector is shown in Table
1. The source fraction of the total counts in the source-extraction
region for each detector is about 70 per cent for the SIS data, 60 per
cent for the first GIS data, and 68 per cent for the second GIS data.
We find no obvious change in observed flux during each observation.
Any possible flux change is less than $\pm 30$ per cent during each 
$\sim 40$ ks long observation (net exposure time is $\sim$20 ks in each). 
Differences of count rate between the two GIS observations (Table 1) are
owing to differences of effective area of the photon collecting region
at the given off-axis positions.
The effective area obtained through the XRT and GIS responses 
in the first observation is $\sim 48$ per cent of that in the 
second observation.
This means that there is no significant change in source flux 
between the two observations after the correction.

\begin{table}
\begin{center}
\begin{tabular}{ccccc}
Observation & S0 & S1 & G2 & G3 \\
1st [\cps] & --- & --- & 0.028 & 0.020 \\
2nd [\cps] & 0.050 & 0.038 & 0.055 & 0.048 \\
\end{tabular}
\caption{The background-subtracted count rate in each detector. There is no SIS data in the first observation. NGC4388 is observed at off-axis angle $\sim 
18$ arcmin in the first observation, 
and $\sim 10$ arcmin in the second observation. The effective area of the 
source-photon collected region of NGC4388 in the first observation is
$\sim 48$ per cent of that in the second observation.}
\end{center}
\end{table}

\section{RESULTS}
\subsection{Spectral fitting}

We have two SIS and four GIS spectra. Since the off-axis position of the
source (and thus the effective area of the XRT) is different for the
first and second observations, we treat the two GIS data sets separately,
and fit the four spectra jointly. There is some difference in the
soft X-ray emission detected by the SIS and GIS data. This is probably
due to uncertainties in background subtraction; we include it as a
systematic uncertainty in our subsequent analysis.
Since we believe the background subtraction is not seriously incorrect,
from the detailed inspection discussed in Section 2.3, errors due to background
subtraction in the individual detectors should be smaller than 
the systematic uncertainty, even in the soft X-ray band.

\subsubsection{The Hard X-ray components}

As Fig. 4 and Fig. 5 show, a strong continuum excess is seen above 3 keV
with a sharp emission line around 6.4 keV. This feature is identified
as the K-shell fluorescence line from cold iron. The hard X-ray
spectrum is well fitted by an absorbed power-law and a gaussian line.
The photon-index and absorption column density depend on
the precise spectral model chosen for the soft X-ray component (see later). 

\subsubsection{A Spectral model for the soft X-ray components}

We first examined the SIS data below 3 keV, since the spectral resolution
and detection efficiency for soft X-rays are better in the SIS than the
GIS. A simple power-law does not give a good fit ($\chi^2 = 74.38$ for 41
degrees of freedom). The best-fit photon index is $\Gamma = 1.6\pm 0.4$
with the Galactic absorption \nH = $3\times 10^{20}$\psqcm, which is
consistent with the {\sl ROSAT} PSPC result, $\Gamma\sim$2--3
(Rush \& Malkan 1996). There is a significant line-like excess around 0.8
keV, which could be due to an iron L-shell emission bump. Since the soft
X-ray emission of NGC4388 is spatially extended, as observed by the {\sl ROSAT}
HRI (Matt et al 1994), this feature is probably due to the thermal
emission from ionized gas associated with the galaxy. However, a thermal
spectrum of a temperature implied from the 0.8 keV peak ($kT\sim 0.5$
keV) alone cannot explain all the emission below 3 keV. A power-law fit
to the 1--3 keV data indicates that the spectrum is flat, with $\Gamma =
1.0\pm 0.4$, and inconsistent with the thermal spectrum characterized by
the 0.8 keV peak, which is much steeper in this energy band. Another
component is required to explain the flat spectrum.
If this is another thermal emission component in the galaxy, a temperature
would be much larger than 10 keV.

Possible origins for the flat spectrum are; a) scattered nuclear
continuum, b) part of the transmitted nuclear continuum, and c) a
collection of high-mass X-ray binaries. These will be discussed in
Section 4. In subsequent fits, the component is modelled by a power-law.

Thus a thermal spectrum plus a power-law can be an appropriate model for
the soft X-ray spectrum. We use a Raymond-Smith thermal spectrum assuming
only Galactic absorption, since this extended component may be free from
the intrinsic absorption of the host galaxy. The absorption for the
power-law component remains a free parameter.

\subsubsection{The spectrum of the whole energy band}

The whole spectrum is fitted using the above two soft X-ray components in
addition to the absorbed power-law plus a gaussian. The photon-index of
the soft power-law component (PL$_2$) is not well constrained. Thus it is
assumed to be identical to the hard power-law (PL$_1$). The GIS is less
sensitive to the thermal component because of its poorer spectral
resolution and limited efficiency below 1~keV. Constraints on the
temperature and abundance are worse than those from the SIS data. The
parameters of the thermal component are therefore fixed at the
SIS-derived values, except for the normalization, when fitting the GIS
spectra. The results are shown in Table 2, and the best-fit models for
the both detectors are shown in Fig. 4 and Fig. 5, respectively.

\begin{figure}
\centerline{\psfig{figure=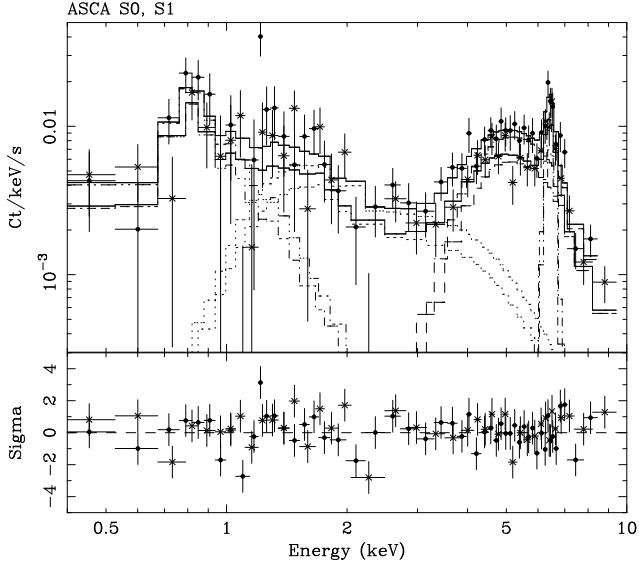,width=0.7\textwidth,angle=270}}
\caption{The {\sl ASCA} SIS spectra of NGC4388. Filled circles are the S0 data and
crosses are the S1 data. The best-fit model in Table 2 is shown in the
stepped-lines. Spectral components are also shown.}
\end{figure}

\begin{figure}
\centerline{\psfig{figure=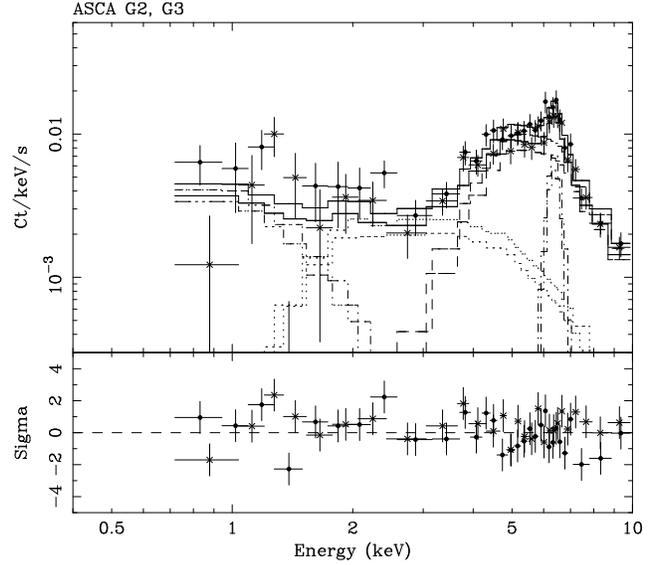,width=0.7\textwidth,angle=270}}
\caption{The {\sl ASCA} GIS spectra of NGC4388. The data from the second observation
only are shown for clarity. Filled circles are G2 data and crosses are G3
data. The best-fit model and spectral components in Table 2 are also
shown.}
\end{figure}

\begin{table*}
\begin{center}
\begin{tabular}{cccccccccccc}
 & \multicolumn{2}{c}{PL$_1$} & \multicolumn{3}{c}{Fe K line} & 
\multicolumn{2}{c}{PL$_2$} & \multicolumn{2}{c}{R-S} & 
$N_2/N_1$ & $\chi^2$/dof \\
 & $\Gamma_1$ & $N_{\rm H1}$ & E & $\sigma$ & EW & $\Gamma_2$ & $N_{\rm H2}$ &
$kT$ & $Z$/\Zs & & \\
 & & $10^{23}$\psqcm & keV & keV & eV & & $10^{22}$\psqcm & keV & & & \\[5pt]
SIS & $1.6^{+0.5}_{-0.4}$ & $4.2^{+0.6}_{-1.0}$ & $6.44^{+0.05}_{-0.05}$ &
$0.14^{+0.13}_{-0.06}$ & $732^{+243}_{-191}$ & $=\Gamma_1$ & 
$1.1^{+0.8}_{-0.9}$ & $0.47^{+0.18}_{-0.17}$ & $0.053^{+0.35}_{-0.03}$ &
0.030 & 93.84/81 \\[5pt]
GIS & $1.5^{+0.4}_{-0.5}$ & $3.7^{+1.0}_{-0.5}$ & $6.39^{+0.06}_{-0.07}$ &
$0.06^{+0.17}_{-0.06}$ & $438^{+127}_{-102}$ & $=\Gamma_1$ & 
$3.1^{+2.1}_{-1.9}$ & $0.47$ & $0.053$ &
0.055 & 219.8/220 \\
\end{tabular}
\caption{Spectral fits to {\sl ASCA} data of NGC4388. The best-fit parameters are
shown for the SIS and the GIS datasets. The first and second data sets
for the GIS were fitted jointly. `PL$_1$' and `PL$_2$' stand for a
power-law for the hard and soft components, respectively, and `R-S' for
the Raymond-Smith thermal spectrum. `$N_2/N_1$' is a normalization ratio
of the PL$_2$ and PL$_1$. These best-fit models are indicated in Fig. 4
and Fig. 5.}
\end{center}
\end{table*}

We detected strong absorption \nH$\approx 4\times 10^{23}$\psqcm ~for the
hard X-ray component, with a relatively small uncertainty compared with
previous measurement (see Table 2). The constraint on the photon index is
weak owing to the limited energy range and strong absorption. The
best-fit $\Gamma\approx 1.6$ is consistent with previous results (Hanson
et al 1990; Takano \& Koyama 1991). A strong fluorescence iron K$\alpha$
line is found for the first time in this object. The equivalent width
(EW) is quite large despite the relatively narrow line width. There is no
evidence for the line broadening in the GIS data whereas there is weak
evidence ($\sigma=0.14^{+0.13}_{-0.06}$ keV) in the SIS data. When we
assume that the PL$_2$ has an identical $\Gamma$ to the PL$_1$, some
extra absorption, \nH$\approx (1-3)\times 10^{22}$
\psqcm, ~is required (see Table 2).
This means that the soft X-ray spectrum around 2 keV is extremely flat.

\subsubsection{X-ray fluxes}

The observed fluxes are summarized in Table 3. The flux in the 3--10 keV
band is slightly smaller than that observed by the SL2~XRT ($f_{\rm 3-10
keV} =2.0\times 10^{-11}$\ergpspsqcm; Hanson et al 1990). The soft X-ray
flux is roughly consistent with previous measurements from the {\sl ROSAT} PSPC
($f_{\rm 0.1-2 keV} = 4.7\times 10^{-13}$\ergpspsqcm, Rush \& Malkan
1996) and the {\sl ROSAT} HRI ($f_{\rm 0.1-2.4 keV} = 5.8\times
10^{-13}$\ergpspsqcm, Matt et al 1994), extrapolating our best-fit model
down to 0.1 keV, although the {\sl ROSAT} flux depends on the assumed model.

\begin{table}
\begin{center}
\begin{tabular}{lcc}
$f_{\rm X}$ & SIS & GIS \\
\hline
\multicolumn{3}{c}{0.5--10 keV [$\times 10^{-11}$\ergpspsqcm]}\\
\hline
Total & 1.6 & 1.4 \\
\hline
\multicolumn{3}{c}{0.5--3 keV [$\times 10^{-13}$\ergpspsqcm]}\\
\hline
Total & 6.4 &  6.5 \\
R-S & 2.9 & 4.4 \\
PL$_2$ & 3.5 & 2.1 \\
\hline
\multicolumn{3}{c}{3--10 keV [$\times 10^{-11}$\ergpspsqcm]}\\
\hline
Total & 1.6 & 1.4 \\
PL$_1$ & 1.3 & 1.1 \\
PL$_2$ & 0.11 & 0.15 \\
\hline
\end{tabular}
\caption{Observed X-ray fluxes in various energy bands. Flux of each
spectral component is also shown, which is derived from the best-fit
model in Table 2. Note that the total flux in the 3--10 keV band include
the iron K line flux.}
\end{center}
\end{table}

\section{DISCUSSION}
\subsection{The obscured nucleus}

We find clear evidence for strong obscuration along the line of sight.
The column density \nH$\approx 4\times 10^{23}$\psqcm ~implied from the
{\sl ASCA} spectrum is similar to those observed in some other classical
Seyfert 2 galaxies (e.g., Awaki et al 1991; Iwasawa et al 1994). The
photon index ($\Gamma\approx 1.6$) is also consistent with that of
Seyfert galaxies. Our best-fit value can be justified by the fact that a
{\sl SIGMA} observation in the 40--150 keV band (Lebrun et al 1992) is in
agreement with an extrapolation of the result of Hanson et al (1990),
unless any extra component, e.g. Compton reflection component, has a
significant contribution at the hard X-ray band (the reflection
contribution should be small if the inner disk is highly inclined to
our line of sight).

The absorption-corrected 2--10 keV flux of the hard power-law component
is $(4.3\pm 0.6)\times 10^{-11}$\ergpspsqcm, corresponding to a luminosity,
$L_{\rm 2-10 keV} = (1.9\pm 0.3)\times 10^{42}$\ergps. This is within the
range of Seyfert 1 nuclei. Colina (1992) predicted that a UV luminosity,
$L_{\rm UV}=1.7\times 10^{43}$\ergps, is required to explain the high
ionization extended nebula. If NGC4388 has the ratio, $L_{\rm X}/L_{\rm
UV}\sim 0.1$, generally appropriate for active galaxies, then the
intrinsic X-ray luminosity of the central source is reasonable. This
agreement may imply that the radiation of the central continuum is not
significantly anisotropic, because the X-ray measurement is for
line-of-sight radiation while the calculation of the UV luminosity is
made for the ionized nebulae extending perpendicular to the
line-of-sight.

The implied heavy obscuration is opaque to the optical/UV continuum from
the central source, unless the dust to gas ratio is unusually low. Even
in the near-infrared band, HeI$\lambda 1.083\mu$m and Pa$\beta\lambda
1.28\mu$m do not show any broad component (Ruiz, Rieke \& Schmidt 1994).
The displacement between the optical nucleus and the radio/10$\mu$m peak
(Stone et al 1988) is likely due to this obscuration effect, and the
true nucleus should be at the radio/10$\mu$m nucleus position. Even
though deficits of HI and CO gas in this galaxy have been reported (Chamaraux,
Balkowski, \& Gerard 1980; Kenney \& Young 1986),
plenty of cold gas remains close to the central X-ray source. This can
form an obscuring torus such as proposed in the unification scheme of
Seyfert galaxies (e.g., Antonucci 1993). The anisotropic radio source
extending perpendicular to the galaxy (Stone et al 1988; Hummel \&
Saikia 1991) and the conical high-ionization regions observed in the
optical band (Pogge 1988; Colina et al 1987) are compatible with the
idea of collimation of radiation by the torus. We note that the
symmetrical axes of the torus and the stellar disk of NGC4388 are aligned
as well as NGC4945 (Iwasawa et al 1993), while they are misaligned in
most cases (e.g., NGC1068, Antonucci \& Miller 1985; NGC5252, Tadhunter
\& Tsvetanov 1989).

An alternative source of the strong X-ray absorption is the interstellar
medium in the nearly edge-on galaxy disk itself. Recent near-infrared
imaging has revealed that this galaxy has a boxy bulge (McLeod \& Rieke
1995).
This and the dust lane crossing in front of the nucleus can be
responsible for some or all of the absorption. In this case, the flat
spectrum between 1 and 3 keV is possibly explained by transmission of
scattered radiation from the central source seen through the
stellar disk. Since the disk is tilted by $\sim$ 18 degrees from being
exactly edge-on, there may be less absorption along these lines of sight,
so making a flat spectrum in the softer band in addition to the strongly
absorbed component.

We note there will be some scattering in the obscuring medium with \nH 
$\sim 4\times 10^{23}$\psqcm ~leading to a hard power-law component in 
the 1--3 keV band (Yaqoob 1996).
This, however, is at least an order of magnitude fainter than the observed 
power-law in that band, unless the absorbing medium is partially ionized.

\subsection{The iron K line}

A strong iron K$\alpha$ emission line is detected at 6.4 keV. The large
equivalent width ($EW=440-730$~eV) of the 6.4~keV line is one of the
signatures of reprocessing in cold matter. Various calculations of X-ray
spectra emerging from a central power-law source surrounded by a torus do
predict a large EW (Awaki et al 1991; Krolik, Madau \& \.Zycki 1994;
Ghisellini, Haardt \& Matt 1994). However, the observed EW is larger than
the value ($EW\sim$~300~eV) expected from the work of Awaki et al (1991)
with
\nH$\sim 4\times 10^{23}$\psqcm, taking an opening angle of the torus
of about 1/3, as deduced from the optical extended nebula (Pogge 1988).
The excess EW could be due to an enhanced iron abundance.

Although the present data cannot provide any meaningful restriction on
the depth of any iron K absorption edge at 7.1 keV, and hence on the iron
abundance, a deep edge due to a supersolar abundance of iron could
partly explain the flat hard X-ray spectra, found in some {\sl Ginga}
Seyfert 2 spectra (Awaki et al 1991; Smith \& Done 1996),
especially when the column density exceeds \nH$\sim 10^{23}$\psqcm.

A similar iron K line feature has been found in {\sl ASCA} spectra of another
Seyfert 2 galaxy Mrk~3 (Iwasawa et al 1994). Mrk~3 showed a significant
intensity change of the iron K line in response to the continuum
variation between two observations of 3.6 yr apart. This fact suggests
that the line is produced within $\sim$1 pc from the central source. We
do not of course detect any variability within the present one-day
observing run of NGC4388 and therefore cannot obtain any constraint about
the line emitting region from this observation. 

\subsection{The scattered continuum}

There is a flat spectral component detected in the intermediate energy
range between the hard component and the thermal component. Such a flat
component has been observed in {\sl ASCA} spectra of Seyfert 2 galaxies NGC1068
(Ueno et al 1994) and NGC6552 (Fukazawa et al 1994), and the eclipse
spectrum of X-ray binary Vela X-1 (Nagase et al 1994). These are
considered to be scattered continuum of the central source. It is
therefore plausible that this component in NGC4388 is also scattered
radiation from the central obscured source, although the possibility
that it is radiation from the central source escaping through a partially
covering absorber as mentioned in Section 3.1.2. is also viable.

As the broad H$\alpha$ detected from the off-nuclear region (Shields \&
Filippenko 1988; 1996) suggests, scattered light may be observed in the optical
band. However, spectropolarimetry in the optical blue band does not
detect any such broad-line region (Kay 1994).

One can expect that several emission lines produced in the photoionized
gas or cold gas which is responsible for the scattering whould be
observable on the X-ray continuum, as seen in the {\sl ASCA} spectra of Mrk~3
(Iwasawa et al 1994) and NGC6552 (Fukazawa et al 1994; Reynolds et al
1994). It is, however, hard to detect such lines in our data, because of
poorer statistics and contamination by the thermal emission. The
detected absorption column density of a few times $10^{22}$\psqcm ~imposed on 
the power-law is consistent with the narrow-line region reddening
E(B--V)$\sim 0.5$ mag deduced from the [SII] doublet (Malkan 1983),
although a recent estimate of the reddening from 
Pa$\beta$/H$\beta = 1.1$ gives a greater value, 
E(B--V)$\sim 1$ mag than this (Ruiz et al 1994). 
We do however detect a marginally significant iron line due to
hydrogenic iron in the SIS spectrum at $6.9\pm0.1$~keV of equivalent
width $194\pm 118$~eV against the absorbed hard continuum or
$4.7\pm2.9$~keV against an extrapolation of the scatttered continuum. The
strength of this line is difficult to predict as it is probably due to
resonant scattering of the continuum (Matt, Brandt \& Fabian 1996), but
it does corroborate the scattering hypothesis. To be so highly ionized,
the scattering medium is probably within a pc of the nucleus for the
observed X-ray luminosity of the central source.

We note that a reasonable fit to the whole SIS spectrum can be obtained with a
partially-covered power-law spectrum, with a gaussian line at 6.4~keV and
an edge at 0.9~keV. This last feature could be due to a oxygen warm
absorber, such as is commonly observed in the spectra of Seyfert 1
galaxies, along the line of sight to the scattering medium. 
This model is rejected because the {\sl ROSAT} HRI image shows that 
the scattering medium is much more extended than can be plausibly ionized 
by the central source (Matt et al 1994).

Other explanations for the flat spectrum are transmission through the
edge-on stellar disk, as discussed in the previous section, and a
population of high mass X-ray binaries. This last possibility is unlikely
given the expected massive star population in the galaxy. The ratio of far
infrared luminosity to blue luminosity can be an indicator of relative
excess of young population (O stars and high mass X-ray binaries; David
et al 1992). Since this ratio in NGC4388, $L_{\rm FIR}/L_{\rm B}=1.5$,
is a factor of 10 lower than in M82, it is unlikely that an intensive
starburst like that in M82 is taking place in NGC4388. A collection of
X-ray binaries is expected to emit a much smaller X-ray luminosity than
that observed (\Lx$\approx 5\times 10^{40}$\ergps).

\subsection{The extended thermal emission}

The {\sl ASCA} spectrum, combined with the extended image (Fig. 1)
resolved by the {\sl ROSAT}
HRI (Matt et al 1994), provides evidence that the soft X-ray emission
below 1keV is of thermal origin. Such thermal emission has been observed
in the {\sl ASCA} spectra of other nearby spiral galaxies which contain an
obscured Seyfert nucleus (e.g., NGC1068, Ueno et al 1994; NGC4258,
Makishima et al 1994). The temperature is $kT = 0.47\pm 0.18$ keV and
the abundance is extremely low ($Z = 0.053^{+0.35}_{-0.03}$\Zs). 
The 0.5--3 keV luminosity of this thermal emission is \Lx = $(1.4\pm 0.3)
\times 10^{40}$\ergps, consistent with the {\sl Einstein} IPC measurement
(Forman et al 1979). Our spectral fits shows most of the observed X-rays
below 1.5 keV come from the thermal component (see Fig. 4 and Fig. 5).
This is compatible with the fact that most of the X-ray emission
detected by the {\sl ROSAT} HRI is extended over the nucleus (Matt et al 1994),
taking account for the {\sl ROSAT} bandpass.

The extended thermal gas could be the stripped interstellar medium of the
galaxy. Starburst activity is not strong in NGC4388, as discussed in
Section 4.3, but consistent with that in normal galaxies with
respect to the $L_{\rm X}/L_{\rm FIR}$ correlation (David et al 1992). 
NGC4388 is known to be located close to the Virgo cluster
core, and to move at high velocity of 1311 \kmps ~relative to the cluster
(Corbin et al 1988) in the deep cluster potential The deficiency of HI
and CO in this galaxy has been explained in terms of ram-pressure
stripping by interaction with the Virgo ICM (Chamaraux et al 1980;
Giovanelli \& Haynes 1983; Kenney \& Young 1986). This stripping mechanism
may also explain the X-ray plume of the nearby elliptical galaxy M86
(Forman et al 1979; White et al 1991; Rangarajan et al 1995) which also
has similar high radial velocity relative to the cluster (they may be
part of an infalling group). The lower temperature and lower abundance in
NGC4388 are different from the X-ray plume of M86, however. High quality
soft X-ray imaging, such as will be available with {\sl AXAF}, are needed to
settle this issue.

Finally, we note that at the radius of the thermal component
(1.4--4.5~kpc, Matt et al 1994) the gas density is only
$\sim0.2$~cm$^{-3}$, assuming bremsstrahlung emission since the abundance
is so low, and any scattered X-ray emission will be negligible. 
The bremsstrahlung emission from any scattering medium at a temperature 
$>10^6$ K would swamp the scattered flux. The
radiative cooling time of the gas is $\sim 10^8$~yr and the recombination
time (of say highly ionized oxygen) is $<10^6$~yr. The flow time at a
velocity of $300$ \kmps ~is $\sim10^7$~yr. The gas may therefore be due to
an outburst in the last ten million yr, or a more general wind from the
nuclear regions. 
The mass flow rate is then about {\it \.M}$\sim 100v_{300}R_2^2$\Ms yr$^{-1}$
where $v_{300}$ is a wind velocity in unit of 300 \kmps ~and $R_2$ is
the radius of the thermal emission region in units of 2 kpc, which is at the
highest
end of the values for superwind galaxies ({\it \.M}$\sim$1--100\Ms yr$^{-1}$, 
Heckman, Armus \& Miley 1991).
This large mass injection is unlikely for NGC4388 since it shows 
no evidence for a strong starburst such as in superwind galaxies.
The low measured abundance (though error is rather large) is also puzzling.
It is however measured mainly for iron-L in our spectrum.
As the abundance of iron is usually lower than that of other 
elements in starburst galaxies, the low abundance may not be unusual if
the thermal emission is produced through a recent starburst.

\subsection{A consistent solution for NGC4388}

We have identified several problems in the interpretation of the X-ray data of
NGC4388: a) the large soft X-ray extent, if thermal, means a large mass loss
rate from the bulge which can only plausibly last a few million yr, b) a 
thermal model requires a very low abundance 
in the extended gas and c) the iron line is
high for the observed obscuration (and in the case of an iron line from
reflection assuming that the solid angle subtended by this and thicker gas at
the nucleus is from a toroidal geometry aligned with the disk of the galaxy). 

These unusual propoerties can be solved
if we assume that the nucleus has not always
had its present luminosity, $L_{\rm x}$, but had a higher luminosity, $L'_{\rm
x}$ a thousand years or more ago when gas at the radius of the soft X-ray
extent ($\sim 1.5$~kpc) was being irradiated. The warm absorber solution in
Section 4.3 requires that the ionization parameter, $\xi=L/nR^2$ where the
density of gas at radius $R$ is $N$, is $\sim 100$. Then  $$L'_{\rm x}=\xi n
R^2=\xi{L_s\over{f\sigma L'_{\rm x}}} R,$$ where $f$ is the fraction of the
volume within $R$ being irradiated and $\sigma$ is the Thomson cross-section.
We then find $L'_{\rm x}\sim f^{-1/2}10^{44}$\ergps, the Thomson depth of the
region is then $\sim f^{-1/2}2\times 10^{-4}$ and the density $\sim
f^{-1/2}0.06$ \pcmcu. 
Provided that $f$ is not small the luminosity is plausible
for a central engine of more than $10^6$\Ms. It is possible that the optical
line emission only shows up one side of a very wide cone so $f>0.5$.

With this solution, all of the X-ray emission below 3~keV can be due to
scattering by partially-ionized gas within a few kpc of the nucleus. The
feature in the spectrum around 1~keV is then due to absorption by OVIII. 
The OVIII absorption edge would be produced in a region close to
the nucleus as found in MCG--6-30-15 (Otani et al 1996). The presence of dust
in the ionized gas implied from the optical/near infrared extinction is 
not peculiar but similar to the dusty warm absorber observed in 
IRAS13349+2438 (Brandt, Fabian \& Pounds 1996) as well as MCG--6-30-15. The
nucleus must have been 100 or more times more luminous in the past. Enhanced
iron K-line emission could be produced by gas in the bulge and disk of the
galaxy (see Fabian 1977).

A continuous decrease in X-ray luminosity over a long time scale has been
observed in some Seyfert galaxies such as NGC~2992 (Weaver et al 1996), 
Mrk~3 (Iwasawa et al 1994) and NGC~1275 (A. Edge, private communication).
The Seyfert 1.9 galaxy NGC2992, for example, has declined in luminosity 
by more than a factor of 20 over the last 17 yr.
The luminosity variation is similar to that found for those Seyfert galaxies,
except that in NGC4388 we require a larger
variation over a longer timescale.
As discussed for Mrk~3 (Iwasawa et al 1994) and NGC~2992 (Weaver et al 1996),
a reverberation effect makes the EW of the iron K line large if the 
X-ray luminosity decreases continuously.

The main problem is the lack of observed optical polarization. This could
however be low due to the small extraction window used by Kay (1994), which
mainly sampled the region obscured by the galaxy disk, and to a wide opening
angle for the emerging radiation, which reduces the predicted level of
polarization. Further spectropolarimetric observations of a wider region are
required.

\section{Summary}

We find from the {\sl ASCA} observations that the spectrum of the Seyfert 2
galaxy NGC4388 has many components. The hard X-ray emission is strongly
absorbed, and a strong iron K line ($EW\approx$ 440--730 eV) is found at
6.4 keV. The hard continuum has a photon index $\Gamma\approx 1.6$ and is
absorbed by a column density
\nH$\approx 4\times 10^{23}$\psqcm. 
Evidence for a thermal spectrum is found in the
soft X-ray band, which can be identified with the spatially extended
emission observed by {\sl ROSAT} HRI (Matt et al 1994). A flat spectrum joins
these two components in the band around 2 keV. It may be the scattered
continuum of the obscured central source.
However, several difficulties still remain; 
(1) an extremely low abundance is required for the thermal emission model; 
(2) an unlikely  high mass flow rate is required to explain the soft 
X-ray extent; and (3) the iron-K line stronger than expected from the torus 
model, with the observed column density.
These can be overcome if the central source was a hundred times brighter 
a thousand or more years ago. All the soft X-ray emission below 3 keV can
then be due to the continuum scattered in a partially ionized medium 
irradiated by the luminous source. The iron K line produced by gas in
the bulge and disk in the galaxy then remains strong, because of the larger
size. This will be tested by a detailed imaging spectroscopy with future 
missions like {\sl AXAF} and {\sl Astro-E}.

\section*{Acknowledgements}

We thank all the members of the {\sl ASCA} PV team, 
Giorgio Matt and 
David White for useful
discussion and Niel Brandt for helping to make the {\sl ROSAT} HRI image. 
The optical image of NGC4388 was obtained through the {\it Sky View} facility 
operated by HEASARC at NASA/GSFC. 
This research has made use of data obtained 
through the High Energy Astrophysics Science Archive Research Center Online 
Service, provided by the NASA-Goddard Space Flight Center.
ACF and KI thank the Royal Society and PPARC, respectively for support.

\end{document}